\documentclass[num-refs]{wiley-article}

\usepackage{siunitx}
\usepackage{placeins}
\usepackage{url}
\usepackage{rotating}
\usepackage{array}
\usepackage{gensymb}

\newcommand{\cor}[1]{\textcolor{blue}{#1}}
\usepackage{ulem}
\newcommand{\remcor}[1]{\sout{#1}}
\newcommand{\rev}[1]{\textsuperscript{\textcolor{blue}{#1}}}

\renewcommand{\cor}[1]{\textcolor{black}{#1}}
\renewcommand{\remcor}[1]{} 
\renewcommand{\rev}[1]{}

\papertype{Original Article}
\paperfield{Submitted to Magnetic Resonance in Medicine}

\title{Combined Diffusion-Relaxometry MRI to Identify Dysfunction in the Human Placenta}

\author[1\authfn{1}]{Paddy J. Slator}
\author[2,3\authfn{1}]{Jana Hutter}
\author[1]{Marco Palombo}
\author[2,3]{Laurence H. Jackson}
\author[4]{Alison Ho}
\author[1]{Eleftheria Panagiotaki}
\author[4]{Lucy C. Chappell}
\author[3]{Mary A. Rutherford}
\author[2,3\authfn{1}]{Joseph V. Hajnal}
\author[1\authfn{1}]{Daniel C. Alexander}

\contrib[\authfn{1}]{Equally contributing authors.}

\contrib{{\color{red}\textbf{Manuscript word count: 3987}}}

\affil[1]{Centre for Medical Image Computing and Department of Computer Science, University College London, London, UK}
\affil[2]{Biomedical Engineering Department, King's College London, London, UK}
\affil[3]{Centre for the Developing Brain, King's College London, London, UK}
\affil[4]{Women's Health Department, King's College London, London, UK}

\corraddress{Paddy J. Slator PhD, Centre for Medical Image Computing, Department of Computer Science, University College London, London, UK}
\corremail{p.slator@ucl.ac.uk}

\fundinginfo{National Institutes of Health (NIH) Human Placenta Project, Grant/Award Number: 1U01HD087202-01; Wellcome Trust (Sir Henry Wellcome Fellowship), Grant/Award Number: 201374/Z/16/Z; EPSRC, Grant/Award Numbers: N018702, M020533, EP/N018702/1;
National Institute for Health Research (NIHR) Professorship (Chappell), Grant/Award Number: RP-2014-05-019;  Tommy’s Charity and Holbeck Charitable Trust}

\runningauthor{Slator et al.}

\begin{document}

\maketitle

\begin{abstract}

\textbf{Purpose}: 
A combined diffusion-relaxometry MR acquisition and analysis pipeline for in-vivo human placenta, which allows for  exploration of coupling between T2* and apparent diffusion coefficient (ADC) measurements in a sub 10 minute scan time.

\noindent\textbf{Methods}:
We present a novel acquisition combining a diffusion prepared spin-echo with subsequent gradient echoes. The placentas of {17} pregnant women were scanned in-vivo, including both healthy controls and participants with various pregnancy complications. We estimate the joint T2*-ADC spectra  using an inverse Laplace transform. 

\noindent\textbf{Results}:
T2*-ADC spectra demonstrate clear quantitative separation between normal and dysfunctional placentas.

\noindent\textbf{Conclusions}:
Combined T2*-diffusivity MRI is promising for assessing fetal and maternal health during pregnancy. 
The T2*-ADC spectrum potentially provides  additional information on tissue microstructure, compared to measuring these two contrasts separately.  
The presented method  is immediately applicable to the study of other organs.

\end{abstract}

\section{Introduction}
\label{sec:Intro}
The placenta provides the vital link between mother and fetus during pregnancy. 
It is implicated in many major pregnancy complications, such as  pre-eclampsia (PE) and fetal growth restriction (FGR) \cite{Brosens2011}. 
PE affects 3-5\% of pregnancies \cite{Mol2016} and is a major cause of maternal and perinatal mortality \cite{Duley2009,Say2014}.
Late onset FGR, defined as that diagnosed after 32 weeks \cite{Gordijn2016}, affects 5-10\% of pregnancies \cite{Figueras2018}. It is strongly associated with stillbirth \cite{Gardosi2013, Bukowski2017}, pre-eclampsia \cite{Srinivas2009}, and late preterm birth \cite{Loftin2010}. 
For all these disorders, it is likely that placental dysfunction occurs before the onset of symptoms.
New techniques for imaging the placenta therefore have the potential to improve prediction, diagnosis, and monitoring of pregnancy complications.

Placental MRI is emerging as a technique with substantial promise to overcome some disadvantages of  ultrasound.
For example, ultrasound parameters of fetal wellbeing are imperfect for determining which fetuses have late-onset FGR and are at greatest risk of adverse perinatal outcome, as opposed to those that are constitutionally small but healthy \cite{MacDonald2015,Figueras2018}.
Assessing the placenta with MRI has the potential to make this distinction. 
Two MRI modalities that show great promise for assessing placental function are T2$^*$ relaxometry - which has the potential to estimate oxygenation levels \cite{Huen2013,Sorensen2013}, and diffusion MRI (dMRI) - which can estimate microstructure and microcirculatory properties \cite{Moore2000a,Derwig2013,Sohlberg2014,Slator2017a}. 

T2$^*$ relaxometry exploits the inherent sensitivity of the transverse relaxation time to the biochemical environment of tissue.
In particular, the paramagnetic properties of haemoglobin mean that the T2* \remcor{relaxation rate}\cor{time constant}\rev{R2.9} can be used as a proxy estimation of oxygenation \cite{Zhao2007}. 
In placental studies, T2* is generally  lower in FGR cases \cite{Sinding2017,Ingram2017,Derwig2013a,Sinding2018}. A typical experiment acquires gradient echo data at several echo times (TE), either in separate or multi-echo scans, and hence estimates the T2* \remcor{relaxation rate}\cor{constant}\rev{R2.9} of the tissue. No diffusion weighting is typically applied to these scans.
Applying diffusion gradients with different strengths (b-value) and directions  provides sensitivity to various microstructural length scales and orientations.  These measurements are usually taken at a fixed TE. 
In the placenta, dMRI has shown promise for discrimination between normal pregnancies and FGR \cite{Moore2008,Bonel2010b,Moore2000a,Sohlberg2015,Derwig2013,Song2017}, and early onset PE \cite{Sohlberg2014}.
However, despite the large number of placental T2* and dMRI studies in the literature, no method has shown sufficient discrimination between healthy pregnancies and those with complications to be introduced into routine clinical practice. Methods which combine multiple distinct measurements may provide a way to overcome this. 
Supporting Information Table \ref{tab:literatureDetails} summarises  T2* and dMRI studies in the placenta to date. 

{T2* and dMRI-derived measures are both influenced by the presence and composition of distinct tissue compartments (or \it{microenvironments}).}
\remcor{Recently, combined diffusion-relaxometry MRI is emerging as a promising technique with the potential for increased sensitivity to these tissue microenvironments}\rev{R2.1}
Diffusion-relaxometry MRI can simultaneously measure multiple MR contrasts; for example by varying both TE and b-value it is possible to probe the multidimensional T2-diffusivity (or T2*-diffusivity) space. 
\remcor{This could provide a more eloquent way of probing microstructure at the subvoxel level.}
\cor{MR experiments dating back to the 1990s have simultaneously measured diffusivity and T2} \cite{VanDusschoten1996, Peled1999, Does2000, Hurlimann2002a, Callaghan2003a}\cor{; such experiments are often categorised  as diffusion-relaxation correlation spectroscopy (DRCOSY)} \cite{Callaghan2003}\rev{R2.1}. 
These \remcor{novel}\rev{R2.1} acquisitions naturally pair with  multidimensional analysis techniques which quantify multiple tissue parameters simultaneously, and therefore have great potential to yield fine-grained information on tissue microstructure.
Such \cor{analysis techniques have been recently applied to}\rev{R2.1} combined diffusion-relaxometry experiments \remcor{have been conducted successfully}\rev{R2.1} in the context of nuclear magnetic resonance (NMR) spectroscopy, improving the ability the distinguish different compartments \cite{Bernin2013,DeAlmeidaMartins2018}. Recent work \cor{applying these techniques to imaging has} \remcor{has extended these techniques to imaging, with}\rev{R2.1} applications in the T1-diffusivity \cite{DeSantis2016}, T2-diffusivity \cite{Kim2017a,Veraart2017}, and T1-T2-diffusivity \cite{Benjamini2017} domains. 
These studies have shown that combining diffusion with other MR contrasts leads to more specific quantification of microscopic tissue compartments.
One recent study demonstrated combined T2-diffusivity in the placenta \cite{Melbourne2018}, with the aim to separate signals from fetal and maternal circulations.

A major disadvantage of previous diffusion-relaxometry experiments are the very long scan times required when varying multiple contrast mechanisms, such as  the TE and diffusion encoding.
\noindent In this paper, we propose a combined acquisition and analysis technique which can estimate the T2*-ADC spectrum within a clinically viable timeframe. 
We apply this novel method in the placenta, an organ where T2* and ADC have both been shown to be informative.
As well as demonstrating simultaneous estimation of T2* and diffusivity parameters within a clinically viable time, we hypothesise that the joint T2*-ADC spectrum will provide additional information compared to the individual measures.

\section{Methods}
\subsection{Acquisition: Integrated T2$^*$-Diffusion sampling}
\begin{figure}[bt]
\centering
  \includegraphics[width=0.7\textwidth]{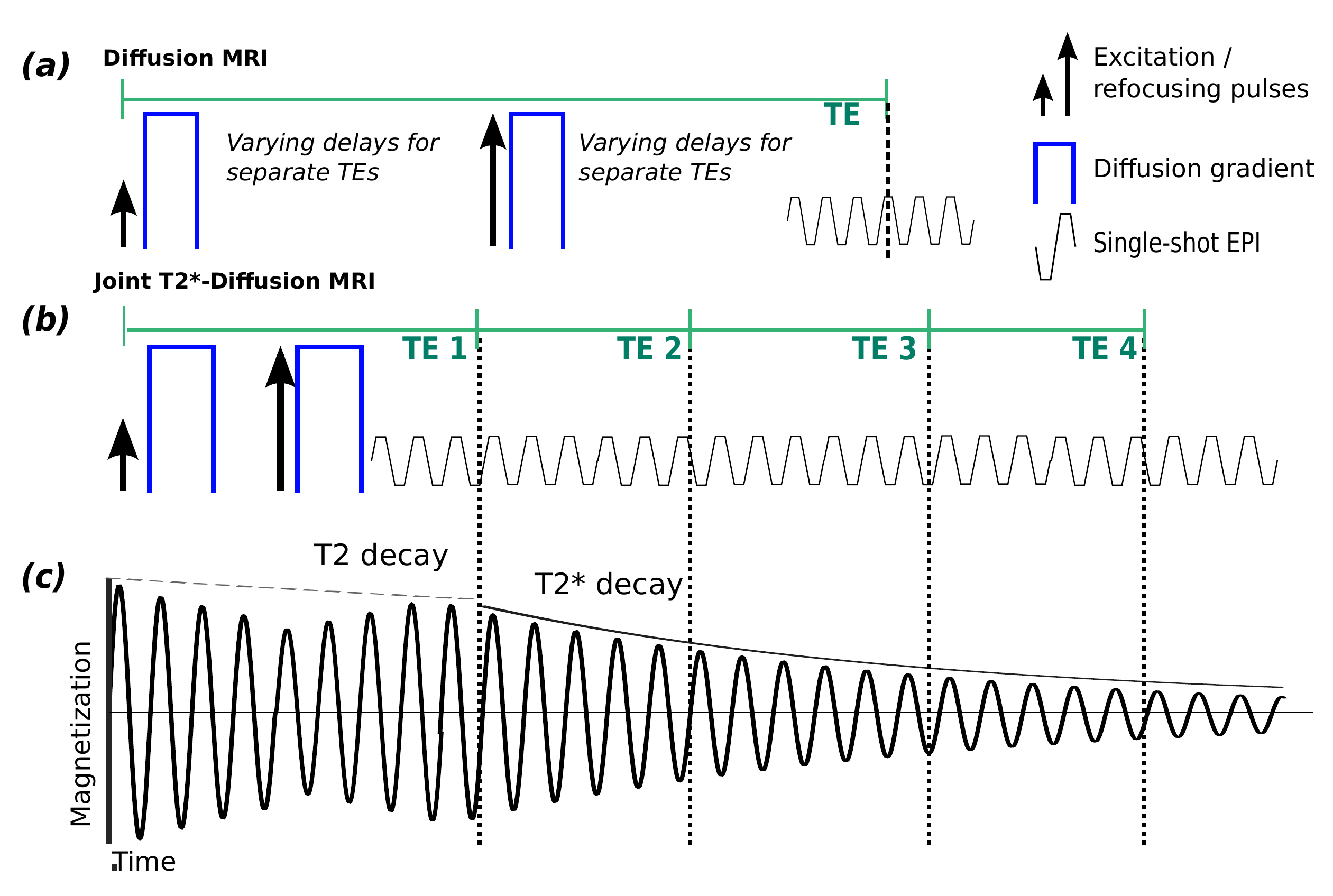}
 \caption{The considered acquisition schemes. (a) Conventional Diffusion MRI acquisition for one echo time (TE) showing the diffusion gradients (blue), the excitation and refocusing pulses as well as the single-shot EPI read-out train. Repeating this acquisition with varying delays between the diffusion gradients and the read-out leads to different TEs and thus combined T2-Diffusion MRI. (b) Proposed combined acquisition with an initial spin-echo acquired after the diffusion gradients followed by multiple Gradient echos. (c) Magnetization for the combined acquisition, with both T2 and T2* decay. The signal evolution neglects effects of all applied gradients.}
   \label{fig:muechos}
\end{figure}

We adapt a novel MRI acquisition strategy, termed ZEBRA \cite{Hutter2018}, in order to sample multiple TEs and diffusion encodings within a single repetition time (TR). The method combines a diffusion prepared spin echo sequence with subsequent gradient echoes. 
This allows simultaneous quantification of T2* and ADC, as opposed to standard independent multi-echo gradient echo and diffusion sequences (e.g. Figure \ref{fig:muechos}a).
Our technique also offers significant speed ups compared to existing T2-diffusivity techniques - which only sample a single TE-diffusion encoding pair for each TR (i.e. Figure \ref{fig:muechos}a).
The proposed combined acquisition is shown in Figure \ref{fig:muechos}b. The multiple  gradient echoes are acquired with minimal spacing after the initial spin echo and diffusion preparation. We note that by using gradient echo readouts rather than spin echoes, we measure  T2$^*$ rather than T2 (see Figure \ref{fig:muechos}c). 

\begin{figure}[t]
  \centering
  \includegraphics[width=0.99\textwidth]{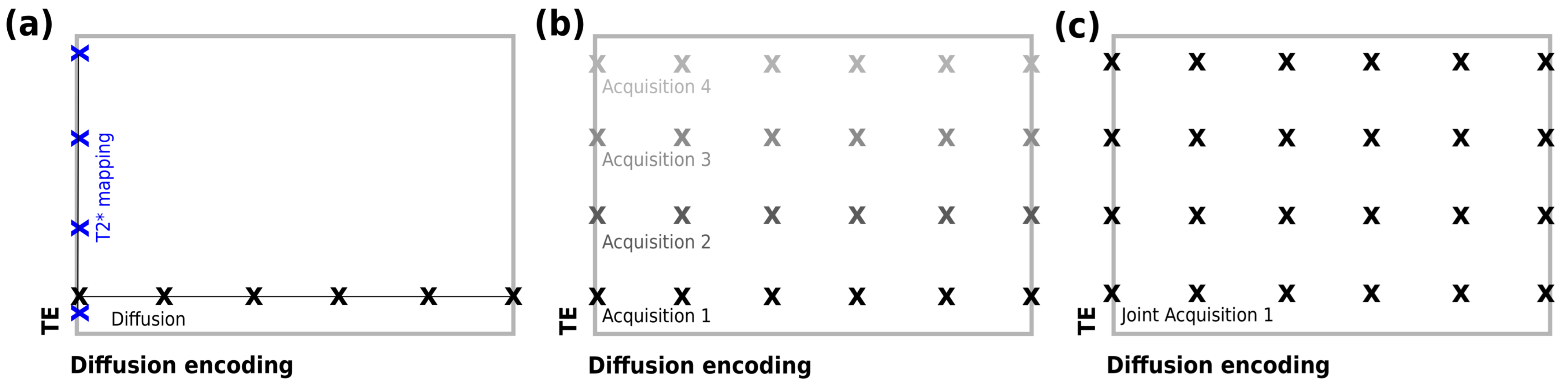}
  \caption{Schemes for the three considered diffusion-relaxometry  experiments illustrated in the TE-Diffusion encoding acquisition parameter plane. (a) Schematic of conventional separate T2* mapping and Diffusion MRI showing the encoding of different echo times for b=0 in blue and different diffusion encoding settings at fixed echo time. (b) Parameter space illustrating the sampling of the TE-diffusivity space with diffusion acquisitions at several TEs. Shading illustrates separate diffusion acquisitions at fixed TEs.
(c) Proposed combined T2$^*$-diffusion acquisition illustrating a denser sampling scheme achieved in a single acquisition.}
  \label{fig:sampling_con}
\end{figure}

Figure \ref{fig:sampling_con} illustrates the resultant sampling of the TE-diffusion encoding domain for the three acquisition techniques presented in Figure \ref{fig:muechos}.
Separate multi-echo gradient echo and diffusion sequences do not adequately sample the full domain (Figure \ref{fig:sampling_con}a).
With repeat acquisitions of diffusion encodings at different TEs  full sampling of the domain is possible, but very slow (Figure \ref{fig:sampling_con}b). The proposed acquisition is able to sample the same domain in a much shorter, and  clinically viable, scanning time (i.e. Figure \ref{fig:sampling_con}c).

\subsection{Modelling}
The simplest model for analysing the data \remcor{assumes}\cor{considers}\rev{R2.2} single tissue compartments, so that the signal attenuations caused by T2* relaxation and diffusion are both \remcor{described by}\cor{assumed to give rise to}\rev{R2.2} a single exponential decay.
The MR signal for this combined ADC-T2* model is given by
\begin{equation}
	\label{eq:t2*-adc}
	S(T_E, b) = S_0  e^{-T_E/T_2^*} e^{-b ADC}
\end{equation}
where $T_E$ is the echo time, $b$ is the b-value, $ADC$ is the apparent diffusion coefficient, $T_2^*$ is the effective transverse relaxation time, and $S_0$ is the signal at the spin-echo time with zero diffusion weighting. 
{$S_0$ is the product of proton density, T2 weighting caused by finite spin echo time, receiver coil properties, and system gain, so we do not treat it as an absolute quantity in the analysis.}

A shortcoming of this model is that it assumes the attenuation due to diffusion is mono-exponential, when it is well established that the placental dMRI signal in-vivo is at least bi-exponential, as in the intravoxel incoherent motion (IVIM) model \cite{LeBihan1988}. In this model, the slow and fast attenuating components are associated with diffusion in tissue and  pseudo-diffusion in capillaries respectively.
Incorporating T2* decay into the IVIM model gives
\begin{equation}
	\label{eq:t2*-ivim}
	S(T_E, b) = S_0  e^{-T_E/T_2^*} \left[ f e^{-bD^*} + (1-f) e^{-b ADC}  \right]
\end{equation}
where $f$ is the perfusion fraction and $D^*$ is the pseudo diffusion coefficient. 
{However, it seems likely that the diffusion and pseudo-diffusion compartments have different T2* values. A model incorporating this was proposed by }
Jerome {\it et al.} \cite{Jerome2016a}
\begin{equation}
	\label{eq:extend-t2*-ivim}
	S(T_E, b) = S_0 \left[ f e^{-bD^*} e^{-T_E/T_{2p}^*} + (1-f) e^{-b ADC} e^{T_E/T_{2}^*}  \right] 
\end{equation}
where $T_{2p}^*$ and $T_{2}^*$ are the T2* values specific to the pseudo-diffusion and diffusion \remcor{compartments}\cor{monoexponential signal components}\rev{R2.2} respectively.

A significant limitation of the models presented in Equations \eqref{eq:t2*-adc} \eqref{eq:t2*-ivim} and \eqref{eq:extend-t2*-ivim} is that the number of \remcor{tissue compartments}\cor{signal components}\rev{R2.2} is assumed to be known.
An alternative approach for analysing the signal is a continuum model, which considers that spins have a spectrum of relaxivity (or diffusivity) values all contributing to the MRI signal. 
Following Menon et al. \cite{Menon1991} the 1D continuum models for $T_2^*$ relaxometry and diffusion are 
\begin{align*}
	S(T_E) &=  \int p(T_2^*) e^{-T_E / T_2^*} \; \mathrm{d} T_2^*	\\
    S(b) &= S_0 \int p(ADC) e^{-b ADC} \; \mathrm{d} ADC.
\end{align*}
Here $p(T_2^*)$ and $p(ADC)$ are the $T_2^*$ relaxation and diffusivity spectra to be estimated from the data. 
We can solve for these spectra using an inverse Laplace transformation, although this is an ill-posed problem requiring regularisation to smooth the resulting spectra \cite{Ronen2006,Bai2014,Ahola2015,Benjamini2017,Kim2017a}.
The extension to combined diffusion-relaxometry acquisitions is simple. For the acquisition presented here, where $T_E$ and $b$ are simultaneously varied, the signal is \cor{(e.g.} \cite{English1991} \cor{) \rev{R2.1}}
\begin{equation}
	S(T_E,b) = S_0 \int_0^{\infty} p(T_2^*, ADC) e^{-TE/T_2^*} e^{-b ADC} \; \mathrm{d}T_2^* \mathrm{d}ADC
\end{equation}
The function we are interested in is the two-dimensional T2*-diffusivity spectrum, $p(T_2^*, ADC)$, which can be estimated by a regularised 2D inverse Laplace transform. 
This contains more information than the individual 1D spectra, and is hence more likely to resolve multiple distinct tissue compartments.
\cor{Although we emphasise that, due to choice of kernels in the continuum models, these distinct compartments - i.e. separate peaks in 2D spectra -  are assumed to be the result of monoexponential signal decays. 
}\rev{R2.2}

\subsection{Experiments}
The sequence described in the methods section was implemented on a clinical Philips Achieva-Tx 3T scanner using the 32ch adult cardiac coil placed around the participant's abdomen for signal reception. All methods were carried out in accordance with relevant guidelines and regulations; the study was approved by the Riverside Research Ethics Committee (REC 14/LO/1169) and informed written consent was obtained prior to imaging. 17 pregnant women, with gestational age ranging from 23+5 to 35+4 (weeks + days), were successfully scanned using the described technique. Three of these participants, one of whom also had FGR, were diagnosed with pre-eclampsia according to standard definitions \cite{Tranquilli2014}.
Three participants had  chronic hypertension in pregnancy and were analysed distinct from normotensive pregnancy women (the control group).
One pregnant woman with chronic hypertension was scanned twice, four weeks apart, and 
developed superimposed pre-eclampsia by the second scan. 
The full participant details are given in Table \ref{tab:patientdetails}.

The combined T2$^*$-diffusivity scan was acquired with the proposed sequence, {a dMRI prepared spin echo followed by multiple gradient echos}. 
The number and timing of the gradient echos varied across scans (see Table \ref{tab:patientdetails}), with most scans having five TEs. The diffusion encodings were chosen specifically for the placenta, as previously reported \cite{Slator2018,Hutter2019}, with 3 diffusion gradient directions at b = [5, 10, 25, 50, 100, 200, 400, 600, 1200, 1600] s mm$^{-2}$, 8 directions at b = 18 s mm$^{-2}$, 7 at b = 36 s mm$^{-2}$, and 15 at b = 800 s mm$^{-2}$. Further parameters were FOV = 300$\times$320$\times$84 mm,  TR = 7 s, SENSE = 2.5, halfscan = 0.6, resolution = 3mm$^3$. One participant was scanned at higher resolution: 2 mm isotropic. The total acquisition time was 8 minutes 30 seconds. 
\cor{We acquired all images coronally to the mother. Attempting to acquire images in the same plane relative to the placenta would be very difficult, due to the heterogeneity in placental positioning and curvature across subjects. In clinical practice the imaging plane with respect to the placenta has to vary widely; our samples allow us to demonstrate the method across a range of orientations.}\rev{R1.3}
\cor{Supporting Information Figure \ref{fig:raw_t2_diff} displays raw data from a single acquisition.}\rev{R1.1}

\subsection{Model fitting}
We first manually defined a region of interest (ROI) containing the whole placenta and adjacent uterine wall section on the first b=0 image with the lowest TE.
We fit the T2*-ADC model described in Equation \eqref{eq:t2*-adc} voxelwise to the data (all TEs and all b-values). 
The fitting consisted of two-step (grid search followed by gradient descent) maximum log-likelihood estimation assuming Rician noise, similar to that previously described \cite{Slator2017a}, with the exception that we use the unnormalised MRI signal. 
The gradient descent fitting constraints were as follows: T2* was constrained between 0.001 s and 1 s, the ADC  between 10$^{-5}$ and 1 mm$^2$ s$^{-1}$, and S0 between 0.001 and $10^5$.
We fixed the SNR for fitting to 20 for all voxels in all scans. 

We calculated the T2*-ADC spectrum for each participant from the signal averaged over the ROIs, using the MERA toolbox \cite{MarkDo}, which incorporates minimum amplitude energy regularization as described by Whittall {\it et al.} \cite{Whittall1989a}.
We also calculated the T2*-ADC spectra voxelwise in all participants. 
We next quantified the spatial variation in T2*-ADC spectral components across the placenta and uterine wall with volume fraction maps, using a similar approach to Benjamini {\it et al.} \cite{Benjamini2017} and Kim {\it et al.} \cite{Kim2017a}.
Specifically, by inspecting the ROI-averaged spectra we chose a set of boundaries  - based on the most common peak areas -  which split the T2*-ADC domain into regions.
These boundaries were the same across all participants, and are 
given in Table \ref{tab:boundaries}.
For each voxel's T2*-ADC spectrum, we then calculated the weight of the voxelwise spectra contained in each of these regions.
By normalising these weights to sum to 1 across all regions, we produced spectral volume fraction estimates for each voxel.
Figure \ref{fig:t2_diff_spectral_maps} shows an illustrative example of this calculation;
the spectral volume fraction essentially quantifies the proportion of each voxel's spectrum which lies in each of the highlighted regions in the top-left panel.

\section{Results}
\begin{figure}[t]
  \centering
  \includegraphics[width=\textwidth]{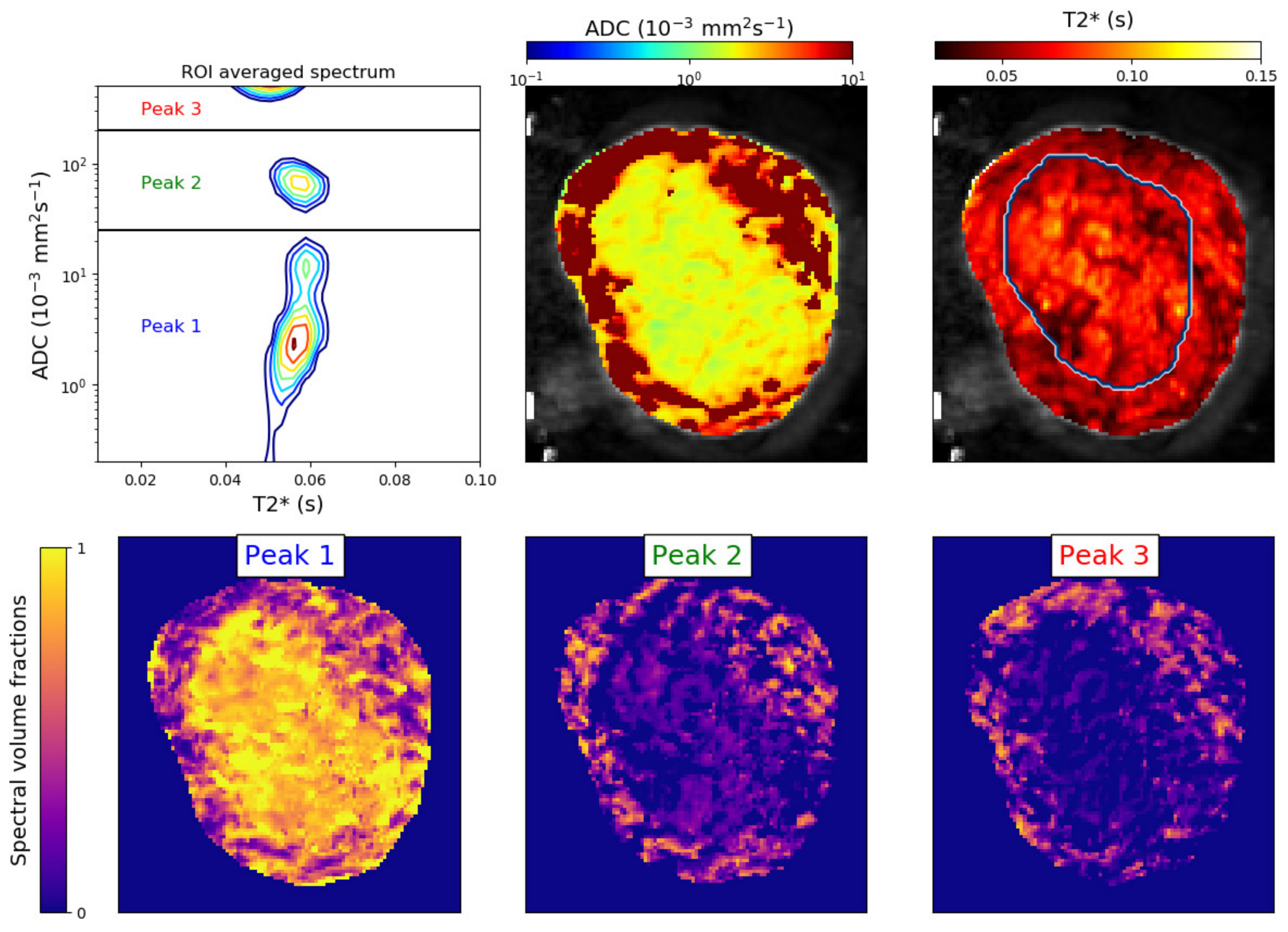}
  \caption{T2*-ADC spectra show anatomical specificity. Spatial maps for a single scan with higher resolution. Top row: T2*-ADC spectrum derived from inverse Laplace transforms of the spatially averaged signal within an ROI comprising the entire placenta and uterine wall, and ADC and T2* maps from combined T2*-ADC fit.
  \cor{The manually-defined placenta ROI is outlined in the T2* map.}\rev{2.10}
  Bottom row, spectral volume fraction maps derived by summing the weight of the spectra in the 3 domains displayed in the ROI averaged spectrum, as described in Methods.}
  \label{fig:t2_diff_spectral_maps}
\end{figure}

\begin{figure}[t]
  \centering
  \scriptsize
  \includegraphics[width=\textwidth]{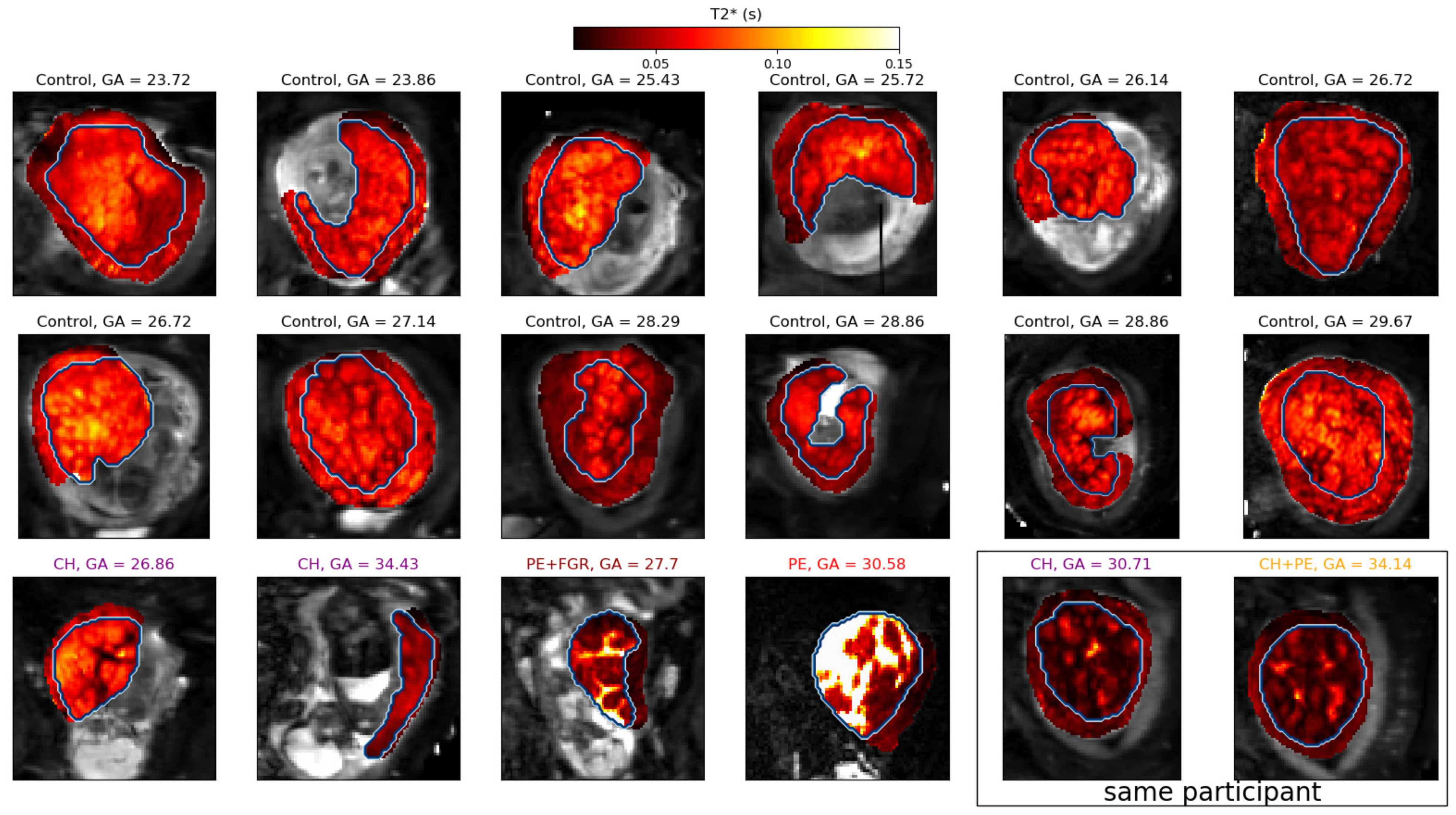}
  \caption{ T2* maps from combined ADC-T2* fit. Participants with pregnancy complications in colour.   \cor{The manually-defined placenta ROI is outlined.}\rev{2.10} \remcor{Note the failure of the model fit in some areas due to very low signal for one PE participant (GA = 30.58).}\cor{Note the very high T2* values for the GA = 30.58 participant - this is very likely due to model fitting failure caused by very low signal in this placenta.}\rev{R2.4}    }
  \label{fig:T2starmaps}
\end{figure}

\begin{figure}[t]
  \centering
  \scriptsize
  \includegraphics[width=\textwidth]{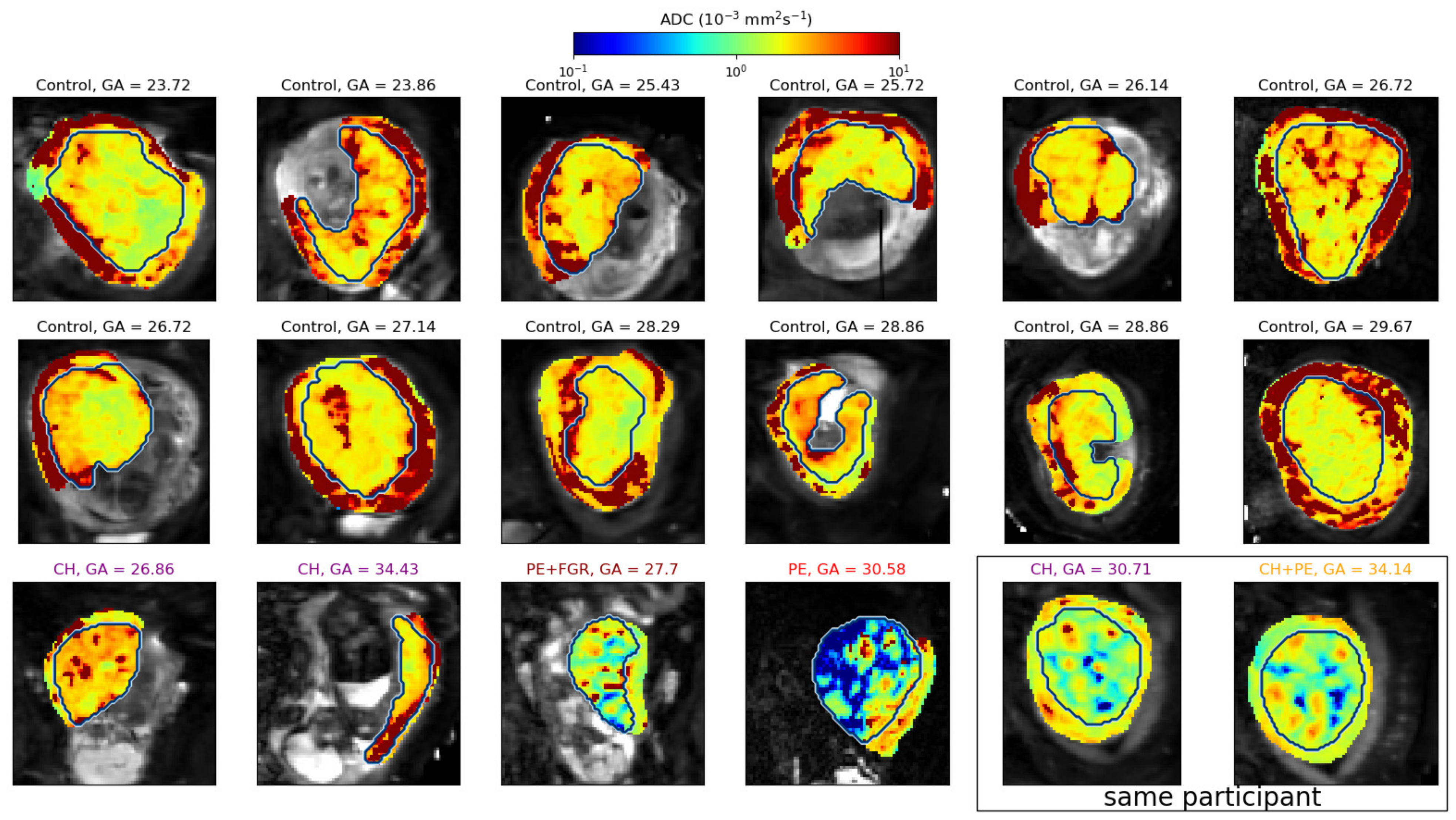}
  \caption{ ADC maps from combined ADC-T2* fit. \cor{The manually-defined placenta ROI is outlined.}\rev{2.10} Note the  log-scale colormap.}
  \label{fig:ADCmaps}
\end{figure}

\begin{figure}[t]
  \centering
  \scriptsize
  \includegraphics[width=\textwidth]{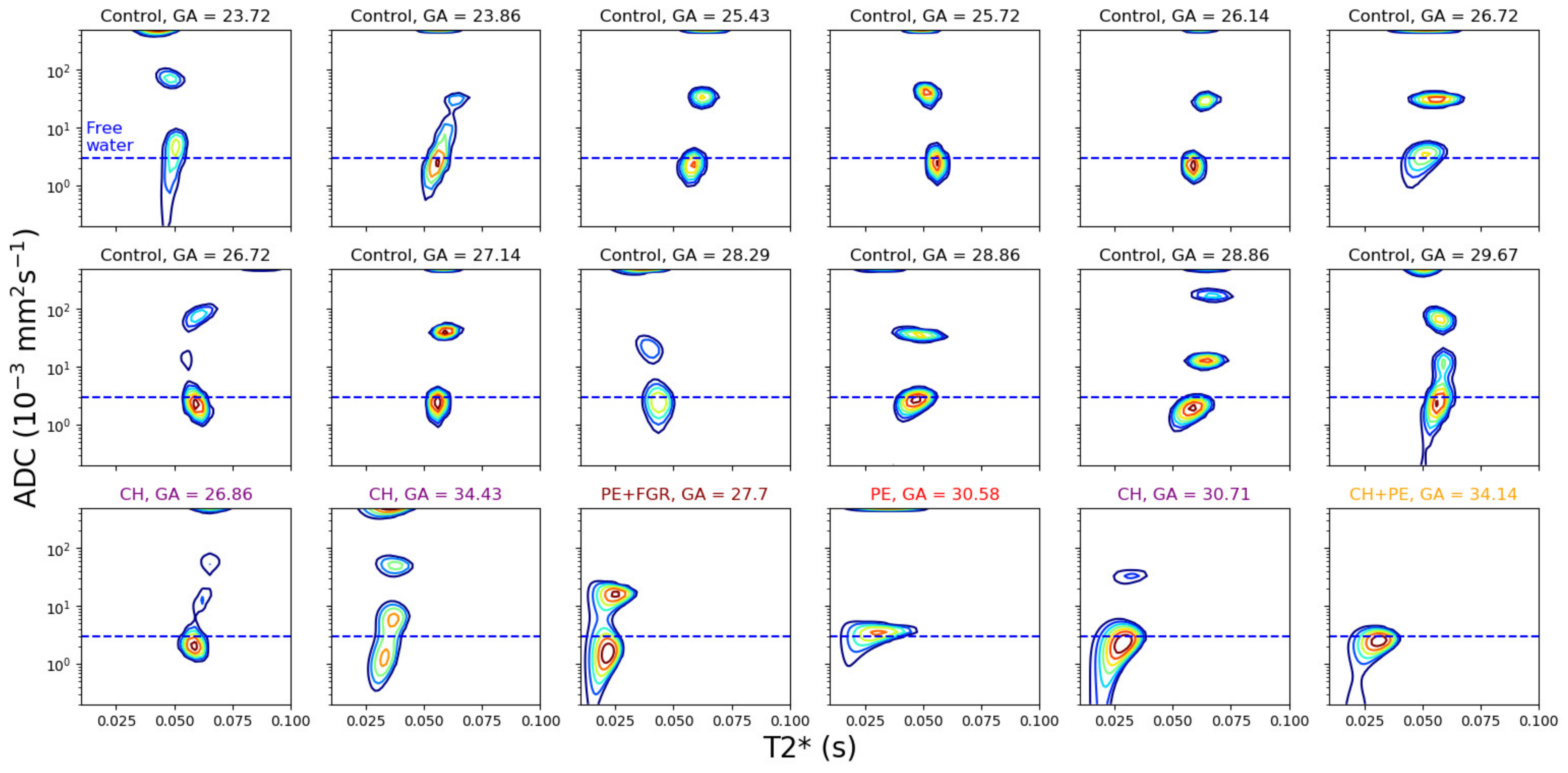}
  \caption{T2*-ADC spectra derived from inverse Laplace transforms of the spatially averaged signal within placenta and uterine wall  ROIs.
    Horizontal dashed blue lines represent the approximate diffusivity of water in free media at 37\degree C (3 $\times 10^{-3} $ mm$^2$ s$^{-1}$).}
  \label{fig:t2_diff_spectra}
\end{figure}

\begin{figure}[t]
  \centering
  \includegraphics[width=0.5\textwidth]{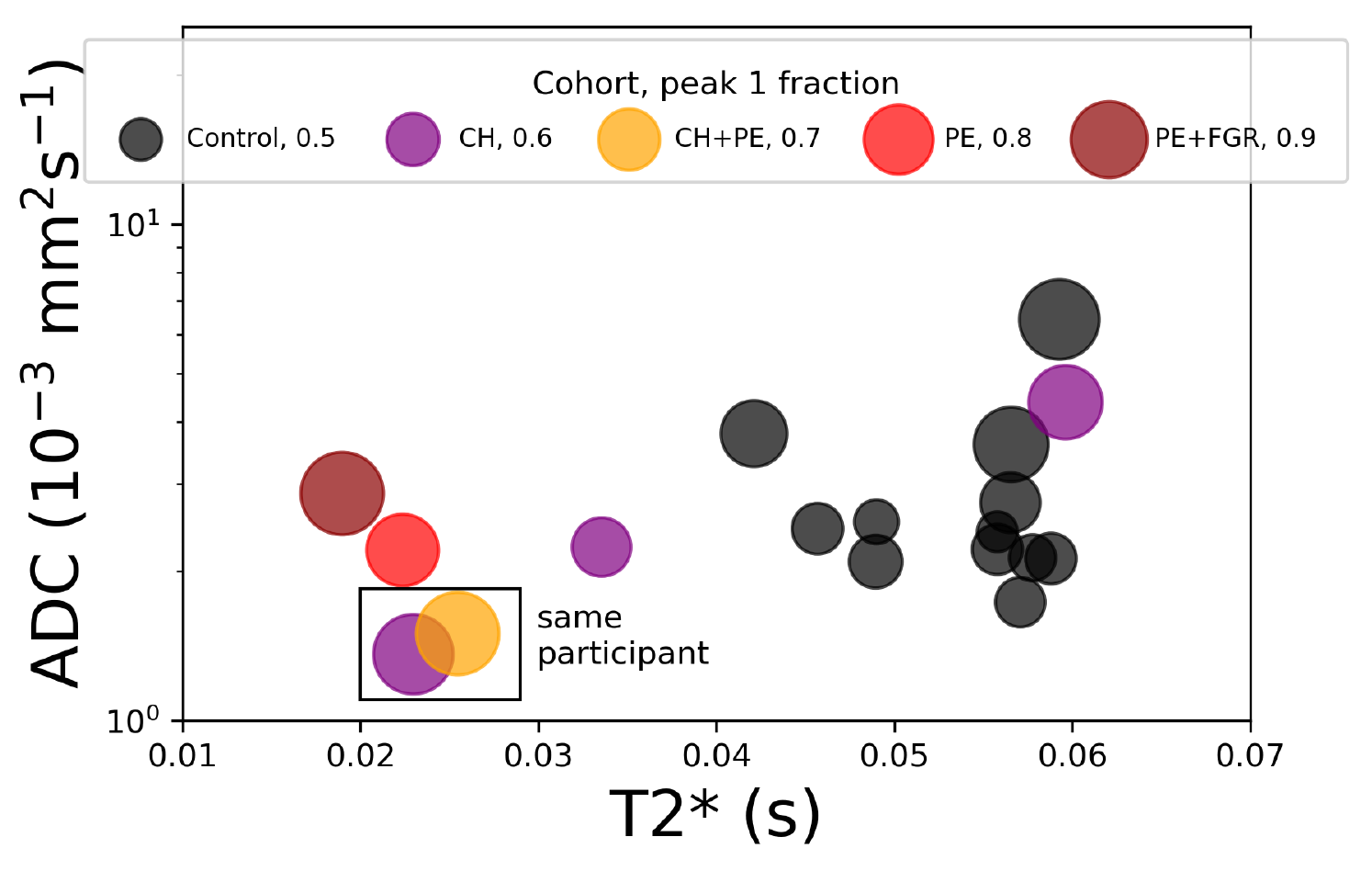}
  \caption{Position of the peak with the lowest ADC within the ADC-T2* spectrum. Each marker corresponds to a single scan. Markers are colored by disease cohort, and marker area is proportional to the spectral volume fraction of the peak. }
  \label{fig:ADC_v_T2}
\end{figure}

Figure \ref{fig:t2_diff_spectral_maps} demonstrates the full analysis pipeline output for a  single participant.
We next present the parameter maps from combined ADC-T2* model fits  (Figures \ref{fig:T2starmaps} and \ref{fig:ADCmaps}) and spectral volume fraction maps (Supporting Information Figures \ref{fig:spectral_maps_1}, \ref{fig:spectral_maps_2}  and \ref{fig:spectral_maps_3}) for all participants.
We probe the changes across gestation and in disease cases by examining the T2*-ADC spectra across all participants (Figures \ref{fig:t2_diff_spectra} and \ref{fig:ADC_v_T2}).
Finally, in order to assess the independence of our diffusivity and relaxometry measurements, we plot the correlation between the derived ADC and T2* values (Supporting Information Figure \ref{fig:adct2starcorr}).

The first panel in Figure \ref{fig:t2_diff_spectral_maps} shows the placenta and uterine wall ROI averaged T2*-ADC spectrum for a single participant (scanned at higher resolution). 
We observe three peaks, clearly separated by ADC value but with similar T2* values. 
ADC and T2* maps show distinctive spatial patterns. The ADC is much higher in the uterine wall than the placenta. T2* maps show distinct `lobes' 
surrounded by a patchwork of low T2* values, with many lobes displaying a small region of higher T2* in the centre.
The bottom row of Figure \ref{fig:t2_diff_spectral_maps} displays voxelwise spectral volume fractions, obtained by integrating (i.e summing spectral weights) within three regions of the T2*-ADC space, as described in Methods. 
The domain with the lowest ADC (e.g. peak 1) is associated with areas within the placenta, and the two domains (peaks 2 and 3) with higher ADC are more prominent in the uterine wall. 

Figure \ref{fig:T2starmaps} shows T2* maps across all participants from the combined T2*-ADC fit.
The patterns are consistent with those previously reported in the literature \cite{Schabel2015,Hutter2019}.
In most participants  regions of high T2* encircled by low T2* borders are clearly visible, and  most likely correspond to placental lobules, with high T2* indicating the presence of oxygenated blood.
In agreement with previous observations  the regions with low T2* are more prominent in pre-eclampsia \cite{Hutter2019}, and FGR \cite{Sinding2018,Lo2017a} placentas.  

ADC maps (Figure \ref{fig:ADCmaps}) also show anatomically-linked  qualitative features which are consistent across participants. 
In all  scans from the healthy pregnant group the ADC shows a significant increase  at the border between the placenta and the uterine wall. This is most likely explained by the high levels of blood flow in these areas.
This bordering area of high ADC is absent from many disease placentas.
Additionally  placentas from women with chronic hypertension and pre-eclampsia often show a distinctive pattern - small patches of high ADC surrounded by very low ADC.


Figure \ref{fig:t2_diff_spectra} displays the spatially averaged T2*-ADC spectra for ROIs containing the placenta and uterine wall.
We clearly observe separate peaks in all control participants, strongly suggesting the presence of multiple tissue compartments with distinct properties. 
In the vast majority (11/12) of these spectra from healthy controls we see at least three clearly separated peaks.
\cor{The ADC values of two of these peaks are typically above the diffusivity of water in free media (Figure \ref{fig:t2_diff_spectra}, blue dashed lines),  suggesting multiple microenvironments with different incoherent flow speeds.}\rev{R2.3}
These peaks, and their corresponding tissue compartments, appear more clearly separated by ADC (note the log-scale on the y-axis) than by T2* value.
We also observed three distinct peaks in placentas from chronic hypertensive women.
Interestingly, we did not see three distinct peaks in any spectra from participants with pregnancy complications (three PE, one PE+FGR). 
There is a distinct pattern in the T2*-ADC spectra for the three PE participants - a left and downward shift in the lowest peak. 
This suggests a decrease in both ADC and T2* distributions compared to control placentas.
There is a similar leftward shift in the PE+FGR placental spectrum; however, the downward shift is not as pronounced, with the middle peak appearing to merge with the lowest peak. 
The peak with highest ADC often appears to span the boundary of the domain in which the inverse Laplace transform is calculated. This is likely because we are unable to sample enough low b-values to accurately estimate this very fast diffusing component - i.e.  there is signal in the $b=0$ volume, which has all attenuated by the  $b=5$ s mm$^{-2}$ volume.

Spectral volume fraction maps showed similar patterns across all control participants (Supporting Information Figures \ref{fig:spectral_maps_1}, \ref{fig:spectral_maps_2} and \ref{fig:spectral_maps_3}); peaks with higher ADC being more prominent in the uterine wall. This likely reflects the high flowing blood volumes in these areas, akin to the maps in Figure \ref{fig:ADCmaps}.

Supporting Information Figure \ref{fig:adct2starcorr} shows that we did not observe a consistent correlation between T2* and ADC values across participants. 
This suggests that we acquire complementary information from these two MR contrasts.
Interestingly, we did not observe the small placental areas with high T2* and high ADC that we saw in previous work \cite{Hutter2019}.

\section{Discussion and Conclusion}
\subsection{Summary}
This study demonstrates accelerated diffusion-relaxometry MRI on the in-vivo human placenta.  Compared to existing approaches, it allows denser, faster, and more flexible sampling of the 2D (TE - diffusion encoding) acquisition space. This in turn allows visualization of the T2*-ADC spectrum, and thus provides  enhanced capacity to separate multiple tissue microenvironments.
The technique was demonstrated on {17} pregnant participants, including {3} scans on placentas clinically assessed as from women with pregnancy complications.
In the following sections, we first putatively associate the observed T2*-diffusivity spectral peaks with distinct placental tissue microenvironments. We then hypothesise as to how the spectral changes observed in cases with complications reflect changes in these tissue microenvironments.
Finally we discuss the clinical potential of the presented technique, which we emphasise is independent of the biological interpretation.

\subsection{Biological interpretation of T2*-diffusivity spectra}
In all controls, we observed  a  peak with high ADC, typically above $10^{-1}$ mm$^2$ s$^{-1}$.
Additionally, in nearly every  control participant (11/12) we observe two further clearly distinct peaks, with ADC  around 2 $\times 10^{-3}$  mm$^2$ s$^{-1}$ for the lower, and between $10^{-2}$ and $10^{-1}$ mm$^2$ s$^{-1}$  for the middle peak (Figure \ref{fig:t2_diff_spectra}). 

The appearance of three  peaks clearly separated by diffusivity in all but one control placenta is consistent with each peak corresponding to a distinct placental tissue microdomain.
Solomon et al. previously reported three placental compartments in mice \cite{Solomon2014}, with these attributed to a slow-diffusing maternal blood compartment, a fetal blood compartment with diffusivity around two orders of magnitude faster, and an intermediate compartment associated with active filtration of fluid across the fetal-maternal barrier.
We therefore speculatively assign tissue compartments to each of these three peaks in healthy control placentas as follows. The compartment with the lowest ADC, which has typical values (2 $\times 10^{-3}$  mm$^2$ s$^{-1}$) comparable to the diffusivity of water in tissue, is associated with \remcor{maternal blood and water within tissue}\cor{water which is not subject to significant incoherent flow effects - this may be within tissue or slow-moving maternal blood}\rev{R2.6}. The highest ADC compartment is associated with perfusing fetal blood, and the intermediate compartment with fluid transitioning between the maternal and fetal circulations \cor{- a significant proportion of which may reside within tissue}\rev{R2.6}. 
This is consistent with the spectral volume fraction maps for the peaks with higher ADC (Supporting Information Figures \ref{fig:spectral_maps_2} and \ref{fig:spectral_maps_3}), which show higher intensity in the vascular areas bordering the placenta.
\cor{The accuracy of these speculative tissue compartment assignments could be tested by comparison with ex-vivo histology. Although such comparisons are notoriously challenging, achieving detailed correspondence would be highly valuable.}\rev{R1.4, R2.7}

\subsection{Spectral changes in disease}
We observed three main trends in the T2*-diffusivity spectrum which discriminated between control and placentas from women with pregnancy complications:
\begin{enumerate}
	\item The disappearance of one (or both) of the middle and higher peaks
    \item \remcor{A leftwards shift in} The lowest peak \cor{has a lower T2*}\rev{R2.11}
    \item \remcor{A downwards shift in} The lowest peak \cor{has a lower diffusivity}\rev{R2.11}
\end{enumerate}

In  placentas from women with pre-eclampsia we generally saw all three trends (Figure \ref{fig:t2_diff_spectra}).   The \remcor{leftward shift}\cor{lower T2*}\rev{R2.11} mirrors the previously reported decrease in T2* in pre-eclampsia placentas \cite{Hutter2019}.
We saw the same \remcor{leftward shift}\cor{trend}\rev{R2.11} in the FGR+PE case, and note that lower T2* values have also been observed  in FGR placentas \cite{Sinding2016,Sinding2018}.
Regarding the \remcor{downward shift}\cor{lower diffusivity}\rev{R2.11} in the lowest peak, our initial speculation is that \remcor{lower diffusivity}\cor{this}\rev{R2.11} could reflect increased water restriction due to  inflammation - since placental  inflammation is associated with PE \cite{Kim2015}.
This may relate to the disappearance of the middle peak, which we hypothesis could reflect decreased maternal-fetal fluid exchange. Inflammation is a potential mechanism facilitating the reduction in exchange,
\cor{although we emphasise that this speculative link can only be confirmed (or refuted) by comparison with post-delivery histology.}\rev{R2.7}
\noindent Figure \ref{fig:ADC_v_T2} presents these observed changes in the T2*-ADC spectrum in a single plot, showing clear  separation between the control and pregnancy complication (i.e. PE, PE+FGR) participants.
We plot the position of the spectral peak with the lowest ADC in the T2*-ADC domain, with the marker area corresponding to the peak's volume fraction. 
In this way, we capture both the peak shift, and the higher volume fraction due to the disappearance of the middle or higher peaks. 
Although these results are highly encouraging, we clearly need to scan many more participants, both control and women with pregnancy complications, to determine the discriminative power of these measures.

\subsection*{Limitations and Future Work}
We used an ``out-of-the-box" inverse Laplace transform toolbox to calculate the T2*-ADC spectrum. There are a number of known weaknesses for this method, including the need for regularization. 
In this study we chose minimum amplitude energy regularization.
Future work could assess the utility of alternative optimization approaches, such as spatially constrained \cite{Kim2017a}, or constrained by the 1D spectra \cite{Benjamini2017}.

Our T2* estimates are generally lower than those previously reported \cite{Hutter2019}. This may be due to the larger voxel size, leading to partial volume effects around areas with high T2*, such as spiral artery inlets.
It could also be due to signal attenuation due to diffusion during the gradient echoes, something which we did not account for in our analysis.

The presented T2*-ADC spectral analysis assesses the data in two dimensions, but there are more dimensions to the data - such as diffusion gradient direction - which we did not include in our analysis. Therefore this dataset has the potential to be further analysed, for example with microstructural models that account for anisotropy in the signal. 

In this study, we used  b-values and gradient directions optimised for dMRI at a single TE \cite{Slator2018a,Hutter2019}, and the TEs were constrained by the EPI read-out train length. 
Separate optimisation of T2$^*$ relaxometry and dMRI acquisition parameters is  1D (choice of TEs, choice of b-values). However, when moving to combined T2$^*$-diffusion this becomes a 2D problem - for example, in the isotropic case we need to choose optimal TE-diffusion encoding pairs. 
In future, we plan to optimise these TE-diffusion encoding values in order to give the best sampling of the 2D parameter space, and enhance estimation of the 2D spectra. 

\remcor{We manually segmented whole placenta and uterine wall ROIs - a time-consuming step - to calculate the T2$^*$-diffusivity spectra. However a single within-placenta ROI, such as the one defined during our scans in order to aid shimming, may be sufficient to discriminate control and disease cases. This would speed up data processing, and also remove the difficulties when segmenting poorly functioning placentas, which often have little functional tissue. }
\rev{R2.8}

\subsection{Outlook and clinical application}
The combined acquisition and analysis technique presented here offers
fast, simultaneous, and multidimensional assessment of placental T2* and diffusivity in less than 10 minutes.
These two MR contrasts have been shown elsewhere to be sensitive to placental pathologies, we hypothesise that their simultaneous assessment could enable better separation of healthy and poorly functioning placentas.
This is supported by the fact that we did not see consistent correlation between T2* and ADC values (Supporting Information Figure \ref{fig:adct2starcorr}), suggesting that these modalities offer complementary information.
This reinforces the value of the novel technique presented here as a quantitative tool for assessment of pregnancy complications, with the potential to ultimately  inform clinical decisions.
Furthermore, we believe that fast calculation of the T2*-ADC spectrum has many potential applications in other areas of biomedical research.

\section*{acknowledgements}
We thank the midwives, obstetricians and radiographers who played a key role in obtaining the data sets.
We would also like to thank all participating mothers.

\section*{conflict of interest}
The authors have no conflicts of interest to declare.


\bibliography{t2sdiff}

\newpage
\section{Figures and Tables}

\FloatBarrier

\begin{table}[bt]
\caption{Participant details. PE - pre-eclampsia, CH - chronic hypertensive, FGR - fetal growth restriction. }
\label{tab:patientdetails}
\begin{tabular}{ | l | l | l | l | }
\hline
	Participant ID & GA at scan (weeks) & Cohort & TEs (ms) \\ \hline
	1 & 23.72 & Control & 78, 114, 150, 186, 222 \\ \hline
	2 & 23.86 & Control & 78, 114, 150, 186, 222 \\ \hline
	3 & 25.43 & Control & 78, 114, 150, 186, 222 \\ \hline
	4 & 25.72 & Control & 78, 114, 150, 186, 222 \\ \hline
	5 & 26.14 & Control & 78, 114, 150, 186, 222 \\ \hline
	6 & 26.72 & Control & 78, 114, 150, 186 \\ \hline
	7 & 26.72 & Control & 78, 114, 150, 186, 222 \\ \hline
	8 & 27.14 & Control & 78, 114, 150, 186, 222 \\ \hline
	9 & 28.29 & Control & 78, 114, 150, 186, 222 \\ \hline
	10 & 28.86 & Control & 82, 175, 268, 361, 454 \\ \hline
	11 & 28.86 & Control & 78, 114, 150, 186, 222 \\ \hline
	12 & 29.67 & Control & 85, 145, 205, 265, 325 \\ \hline
	13 & 26.86 & CH & 80, 121, 162, 203, 245 \\ \hline
	14 & 34.43 & CH & 78, 114, 150, 186, 222 \\ \hline
	15 & 27.7 & PE+FGR & 78, 114, 150, 186, 222 \\ \hline
	16 & 30.58 & PE & 78, 114, 150 \\ \hline
	17 (scan 1) & 30.71 & CH & 78, 114, 150, 186, 222 \\ \hline
	17 (scan 2) & 34.14 & CH+PE & 78, 114, 150, 186, 222 \\ \hline
\end{tabular}
\end{table}

\begin{table}[bt]
\small
\caption{Boundaries selected to segregate  most common peak areas in T2*-ADC spectra.}
\label{tab:boundaries}
\begin{tabular}{|c|c|c|}
\hline 
\textbf{Region}  &\textbf{ADC Bounds} ($\times 10^{-3}$ mm$^2$ s$^{-1}$) & \textbf{T2* Bounds} (s)
\\
Peak 1 & $ 0 < \mbox{ADC} < 25 $  & $0 < \mbox{T2*} < 0.1$ 
\\
Peak 2 & $25  < \mbox{ADC} < 200 $ & $0 < \mbox{T2*} < 0.1$ 
\\
Peak 3 & $200  < \mbox{ADC} < 1000 $ & $0 < \mbox{T2*} < 0.1$  \\
\hline  
\end{tabular}
\end{table}

\FloatBarrier

\section{Supporting Figures and Tables}
\renewcommand{\thefigure}{S\arabic{figure}}

\renewcommand{\figurename}{Supporting Information Figure }

\def\table{\def\figurename{Supporting Information Table}\figure}
\setcounter{figure}{0}

\newcolumntype{H}{>{\setbox0=\hbox\bgroup}c<{\egroup}@{}}

\begin{table}[b]
\small
\caption{Overview of placental T2$^*$ and dMRI studies to date.}
\label{tab:literatureDetails}
 \begin{tabular}{llllHH}
 \hline 
 \textbf{Reference}  &\textbf{Parameters} & \textbf{Resolution}&\textbf{ROI selection}&\textbf{Cohort}\\
 \hline
 \textbf{T2$^*$}&&&&&\\
 Sinding2016\cite{Sinding2016}& 1.5T, gradient-recalled echo)& 1.37x2.73x8mm&Entire placenta, &24 controls, 4 FGR\\
 &16 TEs(3-67.5)&(2 slices, gap 2mm)&outer border not crossed&\\
 &  BH 12s, 16 controls with repetitions& &\\
 Sinding2017\cite{Sinding2017}& 1.5T, gradient-recalled echo)& 1.37x2.73x8mm&Entire placenta, &67 low risk, 28 moderate,5 high risk\\
 &16 TEs(3-67.5)&3 slices&outer border not crossed&\\
 &  BH 12s&transverse evenly&\\
 Sinding2018\cite{Sinding2018}& 1.5T, gradient-recalled echo)& 1.37x2.73x8mm&entire placenta&17 low BW, 68 normal BW\\
 &16 TEs(3-67.5)&3 planes evenly&adjusted for movements&\\
 &  BH 12s, 16 HC with repetitions& &\\
 Derwig2013a\cite{Derwig2013a}&1.5T, flow-compensated SE (ind. scans)&3.76x3.75x8&representative area of central part&normal growth (minor cong. abn.)\\
 &TEs= 40,80,120,180,240,300,360,440&3 slices, no gap& away from vessels&Ut PI$>$95th, weight $<$10th\\
 Ingram2017 \cite{Ingram2017} &gradient-recalled echo &3.52x3.52&largest contiguous placental region&28 controls, 23 FGR\\
 &5-50ms, 8 sec BH, under O$_2$&1 slices transverse& non-placental tissue removed&\\
 Hutter2018\cite{Hutter2019}  & 2D ss EPI Multi-echo GE &  2x2x2&conservative&50 controls, 2 PE\\
 \hline
 \textbf{dMRI}&&&&&\\
 Moore2000a\cite{Moore2000} & 0.5T, 11 b-values (0-468 s mm$^{-2}$) & 3.5$\times$2.5$\times$7 mm & Entire placenta  & 11 controls \\
 Moore2000b\cite{Moore2000a} & 0.5T,11 b-values (0-468 s mm$^{-2}$)  & 3.5$\times$2.5$\times$7 mm & Entire placenta  & 13 controls, 7 FGR \\
 Derwig2013b\cite{Derwig2013} & 1.5T, 11 b-values (0-500 s mm$^{-2}$) & 3.75$\times$3.75$\times$4 mm & Two: central, whole  & 14 controls, 23 SGA \\
 Sohlberg2015\cite{Sohlberg2015} & 1.5T, 5 b-values (0-800 s mm$^{-2}$) & ??$\times$??$\times$6 mm & excluding artefactual signal loss areas & 19 controls, 11 PE, 3 FGR, 2 FGR+PE \\
 You2017\cite{You2017} & 1.5T, 9 b-values (0-900 s mm$^{-2}$) & 4.38$\times$4.38$\times$4 mm & Entire placenta & 16 controls \\
 Capuani2017\cite{Capuani2017} & 1.5T, 7 b-values (0-1000 s mm$^{-2}$) & 2$\times$2$\times$4 mm & Three: central, peripheral, umbilical & 30 controls \\
 Siauve2017\cite{Siauve2017} & 1.5T, 11 b-values (0-1000 s mm$^{-2}$) & ??$\times$??$\times$5 mm & Three: entire placenta, fetal, maternal & 23, scheduled for termination due to fetal anomalies \\
 Slator2017\cite{Slator2017a}&  3T, 12 b-values (0-2000 s mm$^{-2}$) &2$\times$2$\times$2 mm & Entire placenta & 9 controls\\
 Jakab2017\cite{Jakab2017} & 1.5T and 3T, 17 b-values (0-900 s mm$^{-2}$) & 2$\times$2$\times$4 mm & Central & 33, with a range of non-placental adverse findings  \\
 Hutter2018\cite{Hutter2019}&  3T, 14 b-values (0-1600 s mm$^{-2}$) &2$\times$2$\times$2 mm & Entire placenta & 14 controls, 2 PE
 \\
 \hline  
 \end{tabular}
\end{table}

\begin{figure}[b]
  \centering
  \includegraphics[width=\textwidth]{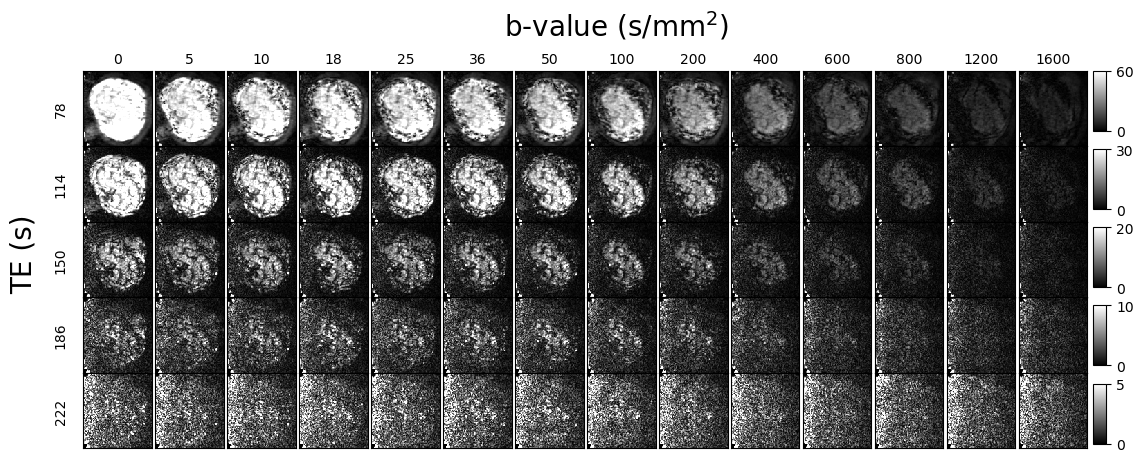}
  \caption{Exemplary raw volumes  from placental diffusivity-relaxometry scan. The resolution was 2 mm isotropic - see Experiments section in Methods for further acquisition parameters.  We display 70 out of the full set of 330 contrast encodings. Note that each row has a different color scaling. Figure \ref{fig:t2_diff_spectral_maps} shows the derived  T2*-ADC spectrum and maps for this scan. }
  \label{fig:raw_t2_diff}
\end{figure}

\begin{figure}[b]
  \centering
  \includegraphics[width=\textwidth]{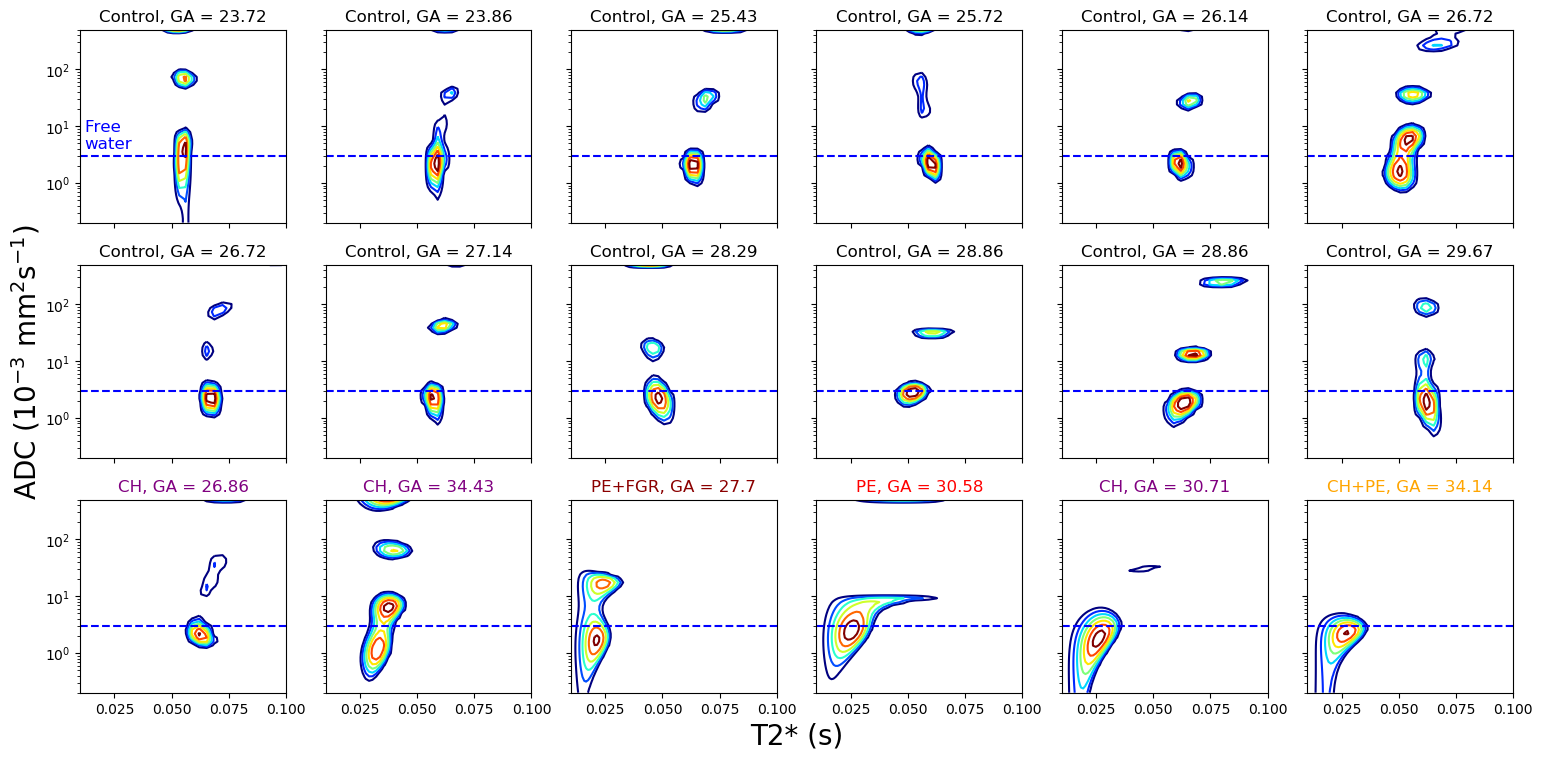}
  \caption{T2*-ADC spectra derived from inverse Laplace transforms of the spatially averaged signal within placenta ROIs.
  Horizontal dashed blue lines represent the approximate diffusivity of water in free media at 37\degree C (3 $\times 10^{-3} $ mm$^2$ s$^{-1}$). }
  \label{fig:t2_diff_spectra_placenta}
\end{figure}

\begin{figure}[b]
  \centering
  \includegraphics[width=\textwidth]{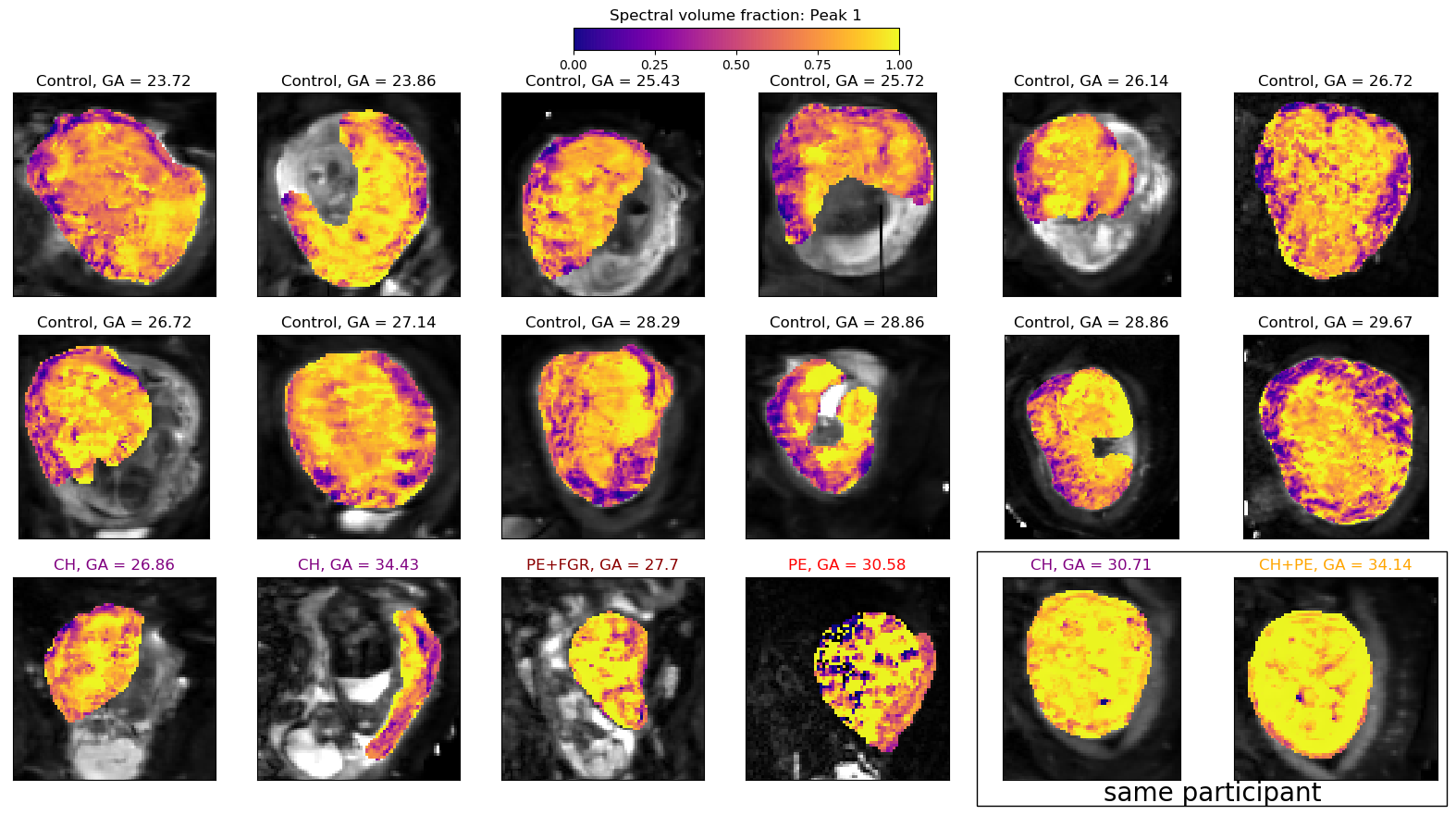}
  \caption{Spectral volume fraction maps, obtained by summing the T2*-ADC spectrum weight  within the domain where $\mbox{ADC} < 25 \times 10^{-3} $ mm$^2$ s$^{-1}$.}
  \label{fig:spectral_maps_1}
\end{figure}

\newpage
\begin{figure}[b]
  \centering
  \includegraphics[width=\textwidth]{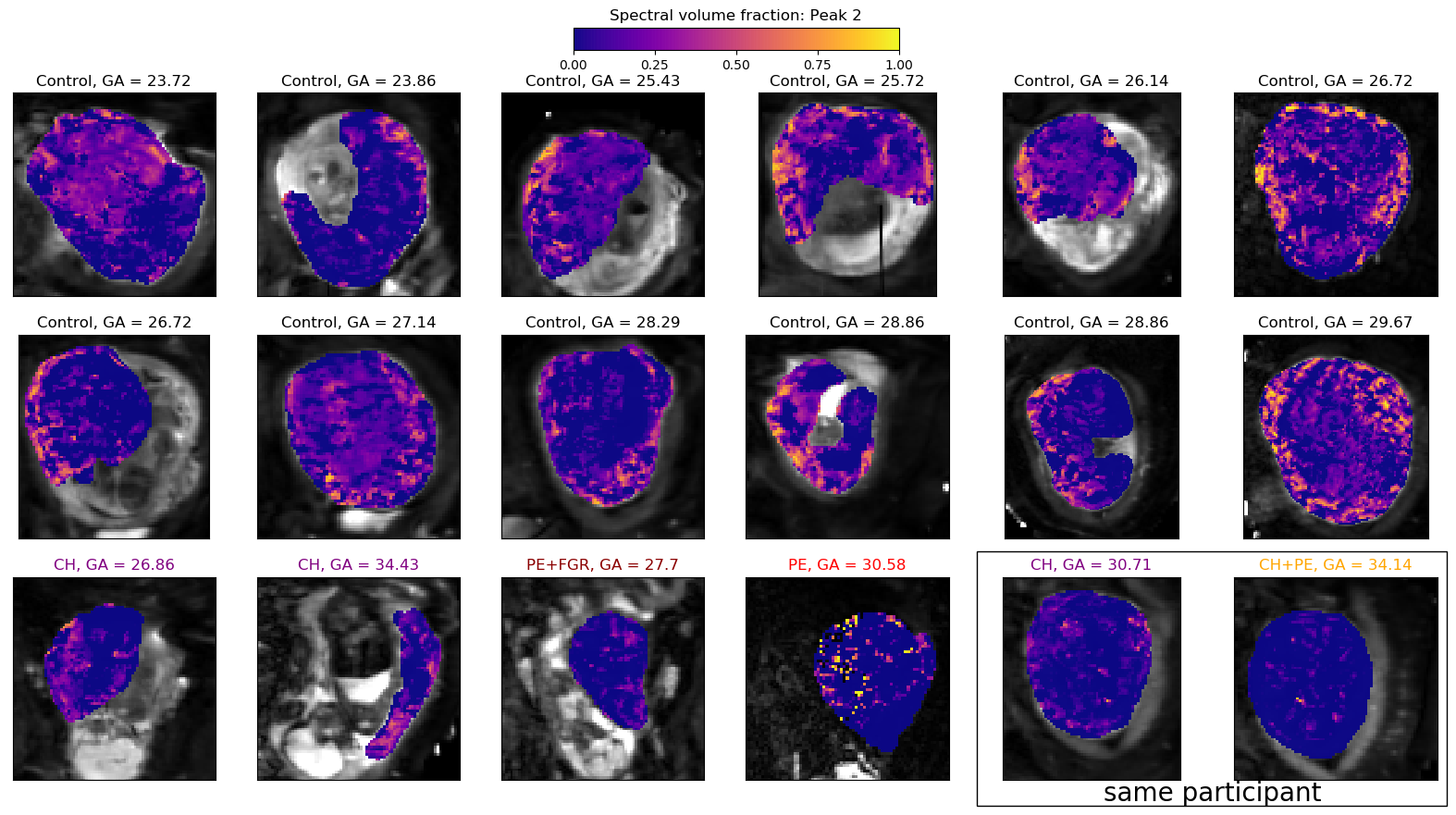}
\caption{As Figure S3, but for the domain where $25 \times 10^{-3}$ mm$^2$ s$^{-1}   < \mbox{ADC} < 200 \times 10^{-3} $ mm$^2$ s$^{-1}$.}
  \label{fig:spectral_maps_2}
\end{figure}

\newpage
\begin{figure}[b]
  \centering
  \includegraphics[width=\textwidth]{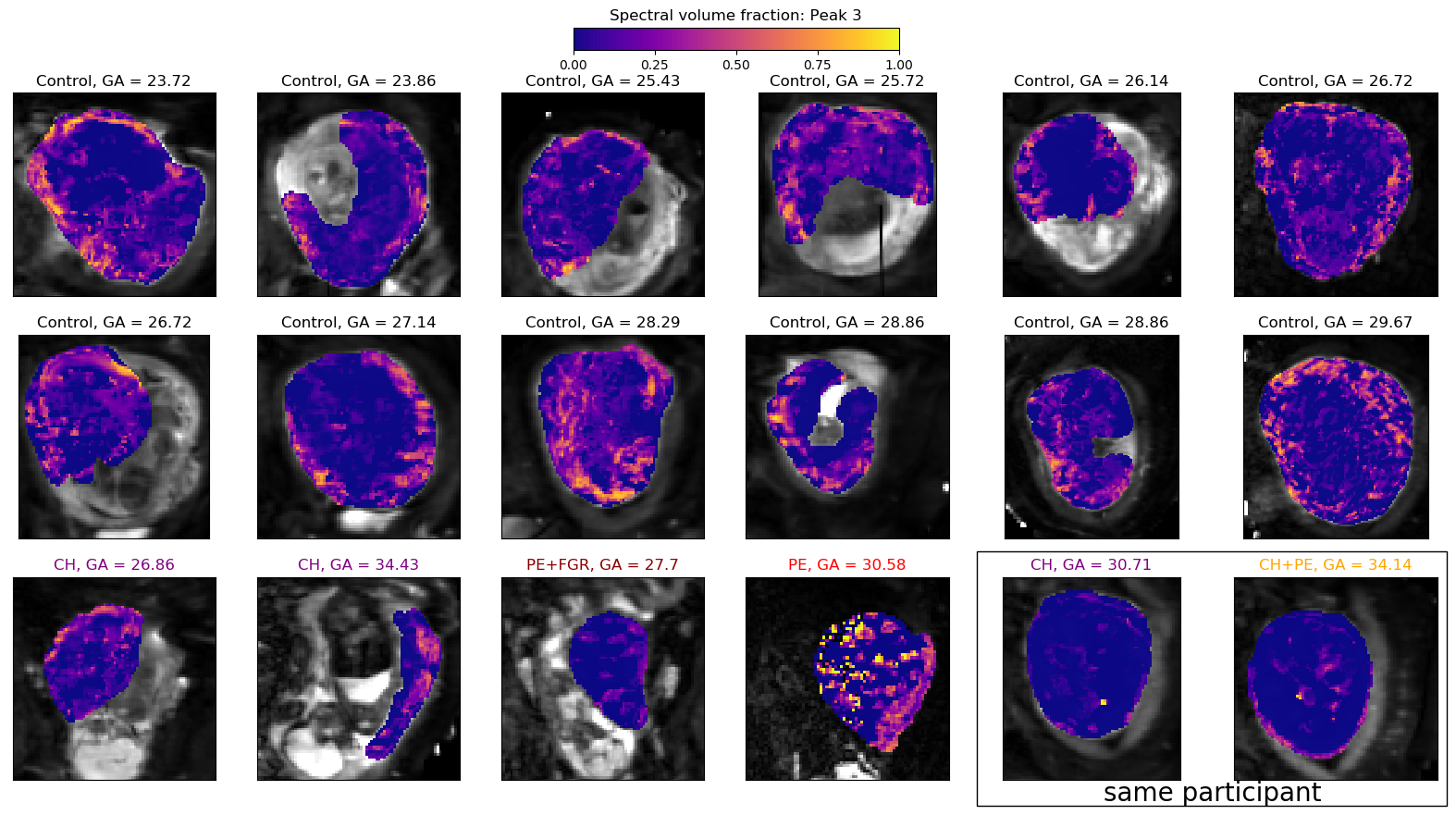}
 \caption{As Figure S3, but for the domain where  $200 \times 10^{-3}$ mm$^2$ s$^{-1}  < \mbox{ADC}  < 1000 \times 10^{-3}$ mm$^2$ s$^{-1}$.}
  \label{fig:spectral_maps_3}
\end{figure}

\begin{figure}[b]
  \centering
  \includegraphics[width=\textwidth]{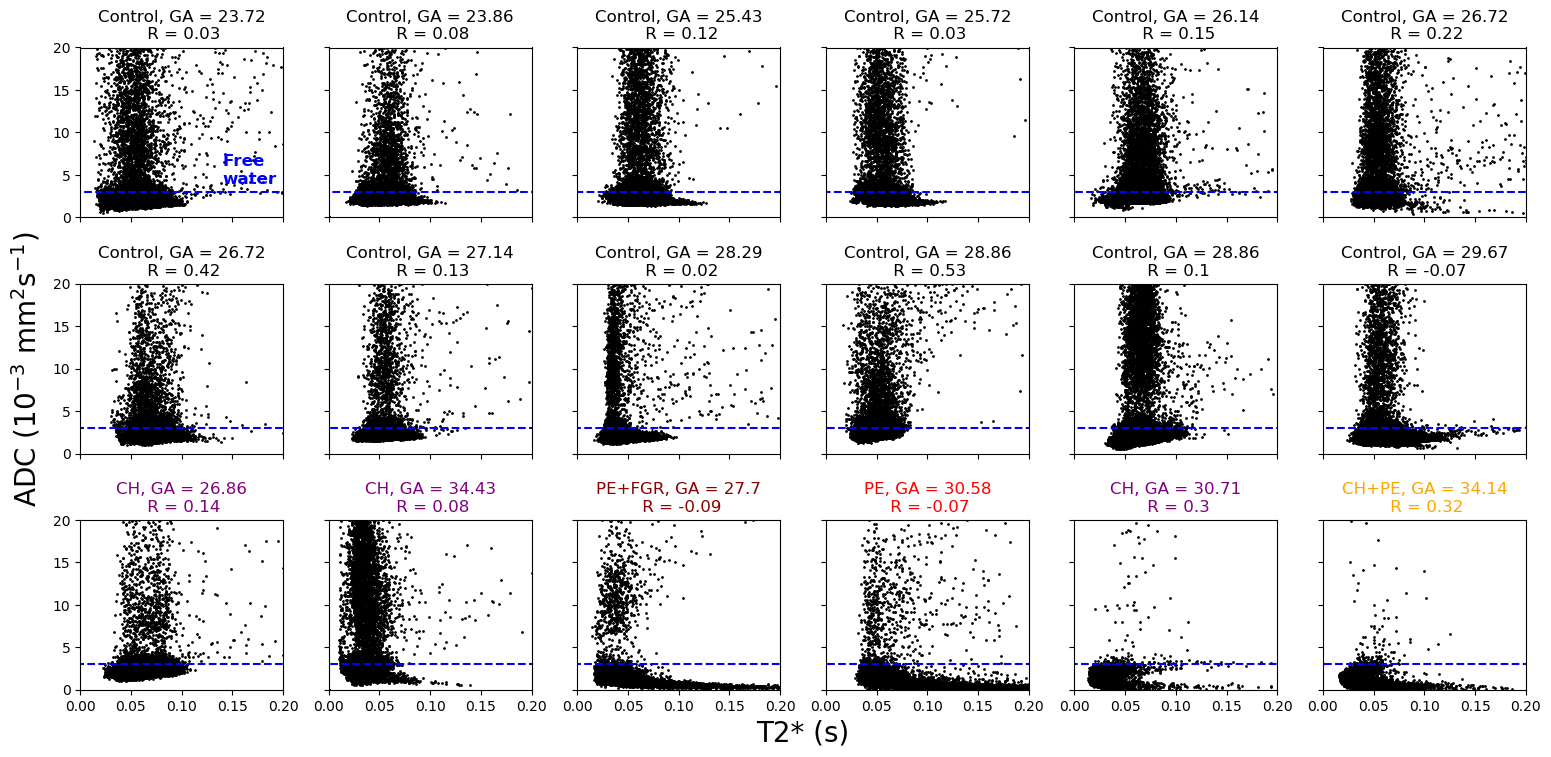}
  \caption{Correlation between T2* and ADC from combined ADC-T2* fit \cor{within placental ROIs}\rev{R2.5}.  Horizontal  blue dashed lines represents the approximate diffusivity of water in free media at 37\degree C (3 $\times 10^{-3} $ mm$^2$ s$^{-1}$). }
  \label{fig:adct2starcorr}
\end{figure}

\end{document}